\documentclass[12pt]{article}
\usepackage{a4}
\usepackage{amsfonts}
\usepackage{amsmath,amssymb}

\makeatother

\def\un#1{\relax\ifmmode\@@underline#1\else
        $\@@underline{\hbox{#1}}$\relax\fi}

\let\du=\du                     

\def\a{\alpha}
\def\b{\beta}

\def\d{\delta}

\def\f{\phi}
\def\g{\gamma}

\def\j{\psi}
\def\k{\kappa}
\def\l{\lambda}
\def\m{\mu}
\def\n{\nu}

\def\s{\sigma}

\def\x{\xi}

\def\F{\Phi}
\def\G{\Gamma}

\def\L{\Lambda}

\def\ve{\varepsilon}
\def\vf{\varphi}

\def\ce{{\cal E}}

\def\cg{{\cal G}}

\def\car{{\cal R}}

\def\cv{{\cal V}}
\def\cw{{\cal W}}

\def\cz{{\cal Z}}

\def\bo{{\raise-.3ex\hbox{\large$\Box$}}}               
\def\pa{\partial}                                       
\def\de{\nabla}                                         
\def\TH{{\raise.2ex\hbox{$\displaystyle \bigodot$}\mskip-4.7mu \llap H \;}}
\def\face{{\raise.2ex\hbox{$\displaystyle \bigodot$}\mskip-2.2mu \llap {$\ddot
        \smile$}}}                                      

\def\abs#1{\left| #1\right|}                    
\def\leftrightarrowfill{$\mathsurround=0pt \mathord\leftarrow \mkern-6mu
        \cleaders\hbox{$\mkern-2mu \mathord- \mkern-2mu$}\hfill
        \mkern-6mu \mathord\rightarrow$}
\def\dvec#1{\vbox{\ialign{##\crcr
        \leftrightarrowfill\crcr\noalign{\kern-1pt\nointerlineskip}
        $\hfil\displaystyle{#1}\hfil$\crcr}}}           
\def\dt#1{{\buildrel {\hbox{\LARGE .}} \over {#1}}}     

\def\frac#1#2{{\textstyle{#1\over\vphantom2\smash{\raise.20ex
        \hbox{$\scriptstyle{#2}$}}}}}                   
\def\sfrac#1#2{{\vphantom1\smash{\lower.5ex\hbox{\small$#1$}}\over
        \vphantom1\smash{\raise.4ex\hbox{\small$#2$}}}} 
\def\bfrac#1#2{{\vphantom1\smash{\lower.5ex\hbox{$#1$}}\over
        \vphantom1\smash{\raise.3ex\hbox{$#2$}}}}       
\def\afrac#1#2{{\vphantom1\smash{\lower.5ex\hbox{$#1$}}\over#2}}    

\def\[{\lfloor{\hskip 0.35pt}\!\!\!\lceil}
\def\]{\rfloor{\hskip 0.35pt}\!\!\!\rceil}
\def\Lag{{\cal L}}
\def\du#1#2{_{#1}{}^{#2}}

\def\fracm#1#2{\hbox{\large{${\frac{{#1}}{{#2}}}$}}}

\def\un{\underline}
\def\fracmm#1#2{{{#1}\over{#2}}}

\def\low#1{{\raise -3pt\hbox{${\hskip 0.75pt}\!_{#1}$}}}

\def\Dot#1{\buildrel{_{_{\hskip 0.01in}\bullet}}\over{#1}}
\def\dt#1{\Dot{#1}}

\def\DDot#1{\buildrel{_{_{\hskip 0.01in}\bullet\bullet}}\over{#1}}
\def\ddt#1{\DDot{#1}}

\newskip\humongous \humongous=0pt plus 1000pt minus 1000pt
\def\caja{\mathsurround=0pt}
\def\eqalign#1{\,\vcenter{\openup2\jot \caja
        \ialign{\strut \hfil$\displaystyle{##}$&$
        \displaystyle{{}##}$\hfil\crcr#1\crcr}}\,}
\newif\ifdtup

\newcommand{\be}{\begin{equation}}
\newcommand{\ee}{\end{equation}}
\newcommand{\nbe}{\begin{equation*}}
\newcommand{\nee}{\end{equation*}}

\newcommand{\lb}{\label}

\begin{document}

\thispagestyle{empty}

{\hbox to\hsize{
\vbox{\noindent January 2009 \hfill UMDEPP 09--023 }}}

\noindent
\vskip2.0cm
\begin{center}

{\large\bf SUPERSTRING--INSPIRED SUPERGRAVITY \vglue.1in
                AS THE UNIVERSAL SOURCE OF \vglue.2in 
                INFLATION AND QUINTESSENCE~\footnote{Supported in part by the 
endowment of the John S.~Toll Professorship, the University of \newline
${~~~~}$ Maryland Center for String \& Particle Theory, US National Science 
Foundation Grant \newline ${~~~~}$ PHY-0354401, and the Japanese Society for 
Promotion of Science (JSPS)}}
\vglue.3in

S. James Gates, Jr.~\footnote{Email: gatess@wam.umd.edu}
and Sergei V. Ketov~\footnote{Permanent address: Department of Physics, Tokyo
Metropolitan University, Hachioji-shi, \newline ${~~~~}$ Tokyo 192--1397, 
Japan. Email: ketov@phys.metro-u.ac.jp}

\vglue.1in
\noindent {\it Physics Department, University of Maryland, College Park, 
MD 20742, USA}
\end{center}
\vglue.2in
\begin{center}
{\Large\bf Abstract}
\end{center}
\vglue.1in

We prove (in superspace) the equivalence between the higher-derivative $N=1$ 
supergravity, defined by a holomorphic function $F$ of the chiral scalar 
curvature superfield, and the standard theory of a chiral scalar superfield 
with a chiral superpotential $W$, coupled to the (minimal) Poincar\'e 
supergravity in four space-time dimensions. The relation between the
 holomorphic functions $F$ and $W$ is found. It can be used as the technical
framework for the possible scenario unifying the early Universe inflation and
the present Universe acceleration. We speculate on the possible origin of our
model as the effective supergravity generated by quantum superstrings, with a
dilaton-axion field as the leading field component of the chiral superfield.  
\vglue.1in
\noindent 

\newpage

\section{Introduction}

Inflation (i.e. a phase of `rapid' accelerated expansion) in the early Universe
 \cite{inf} nicely predicts the homogeneity of our Universe at large scales, 
its large size and entropy, as well as an almost scale-invariant spectrum of
 cosmological perturbations, in good agreement with the high precision CMB 
measurements \cite{pert}. A mechanism (and details) of inflation is usually 
based on a `slow-roll` scalar field (inflaton) with a proper scalar potential 
\cite{inf}. It follows from astronomical observations of Supernova Ia 
\cite{supernova} that the present Universe is accelerating due to the 
mysterious `dark energy' which violates the strong energy condition in General 
Relativity. Dark energy is also needed to prevent a formation of 
super-large clusters of galaxies \cite{vikh}, so it begs for a theoretical 
proof of its existence or an alternative explanation of the present 
acceleration from a fundamental theory of gravity. The most naive explanation 
of the dark energy by a cosmological constant is not satisfactory  because of 
its time-independence and enormous fine-tuning. A better model  is provided by 
a scalar field (quintessence) whose scalar potential is tuned `by hand' 
\cite{quint}. Hence, the true theoretical challenge is to explain the origin 
of inflaton and quintessence, as well as provide theoretical tools for a 
derivation of the scalar potential.
 
The expected scale of inflation is close to that of Grand Unification 
\cite{inf}, so the inflation may be due to some Planck scale physics or 
quantum gravity. A consistent and universal approach to quantum gravity 
and very high-energy particles physics is available due to theory of 
superstrings (or M-theory) \cite{bbs}. Due to its putative fundamental nature,
  string theory is expected to be valid at all energy scales, which would make 
it indispensable in any effort to unify the UV gravity (in the very early  
Universe) with the IR gravity (in the present Universe). String theory should 
also explain the origin of quintessence (or dark energy). Assuming the validity
 of the effective field-theoretical description of string theory, it is 
reasonable to study both inflation and quintessence within the effective 
supergravity framework, because local supersymmetry is required for consistency
 of strings. Of course, supersymmetry is broken in the IR, e.g. spontaneously.

In this Letter we would like to propose a possible {\it geometrical} origin of
the inflaton and quintessence, as described by a {\it single} scalar field in 
the supergravity model modified by higher-order supercurvature terms. The 
latter may originate from quantum (non-perturbative) superstrings, though  we 
do not have a compelling reason for that. We also assume that (i) string theory
 is compactified down to four space-time dimensions, (ii) all of its 
moduli are stabilized (e.g. by fluxes \cite{fluxes}), and (iii) local 
supersymmetry is broken to $N=1$ in uncompactified four dimensions whose 
geometry is described by the FRLW metric with a scale factor $a(t)$ of physical
 time $t$, and $k=(-1,0,+1)$,
\be \lb{frw}
ds_{\rm FRW}^2 = 
 dt^2 - a^2(t)\left[ \fracmm{dr^2}{1-kr^2} +r^2d\Omega^2\right]
\ee

\section{Basic facts of N=1 superfield supergravity}

The chiral superspace density (in the supersymmetric gauge-fixed form) reads 
\be \lb{den}
\ce(x,\theta) = e(x) \left[ 1 -2i\theta\s_a\bar{\j}^a(x) +
\theta^2 B(x)\right]~, \ee
where $e=\sqrt{-\det g_{\m\n}}$, $g_{\m\n}$ is a spacetime metric, 
$\j^a_{\a}=e^a_{\m}\j^{\m}_{\a}$ is a chiral gravitino, $B=S-iP$ is the 
complex scalar auxiliary field. We use the lower case middle greek letters 
$\m,\n,\ldots=0,1,2,3$ for curved spacetime vector indices, the lower case 
early latin letters $a,b,\ldots=0,1,2,3$ for flat (target) space vector 
indices, and the lower case early greek letters $\a,\b,\ldots=1,2$ for chiral
 spinor indices.

The solution of the superspace Bianchi identitiies and the constraints defining
the $N=1$ Poincar\'e-type minimal supergravity results in only three relevant 
superfields $\car$, $\cg_a$ and $\cw_{\a\b\g}$ (as parts of the supertorsion), 
subject to the off-shell relations \cite{ggrw}
\be \lb{bi1}
 \cg_a=\bar{\cg}_a~,\qquad \cw_{\a\b\g}=\cw_{(\a\b\g)}~,\qquad
\bar{\de}_{\dt{\a}}\car=\bar{\de}_{\dt{\a}}\cw_{\a\b\g}=0~,\ee
and
\be \lb{bi2}
 \bar{\de}^{\dt{\a}}\cg_{\a\dt{\a}}=\de_{\a}\car~,\qquad
\de^{\g}\cw_{\a\b\g}=\frac{i}{2}\de\du{\a}{\dt{\a}}\cg_{\b\dt{\a}}+
\frac{i}{2}\de\du{\b}{\dt{\a}}\cg_{\a\dt{\a}}~~,\ee
where $(\de\low{\a},\bar{\de}_{\dt{\a}}.\de_{\a\dt{\a}})$ represent the curved 
superspace $N=1$ supercovariant derivatives, and bars denote complex 
conjugation.

The covariantly chiral complex scalar superfield $\car$ has the scalar 
curvature $R$ as the coefficient at its $\theta^2$ term, the real vector 
superfield $\cg_{\a\dt{\a}}$ has the traceless Ricci tensor, 
$R_{\m\n}+R_{\n\m}-\frac{1}{2}g_{\m\n}R$, as the coefficient at its 
$\theta\s^a\bar{\theta}$ term, whereas the covariantly chiral, complex, 
totally symmetric, fermionic superfield $\cw_{\a\b\g}$ has the Weyl tensor 
$W_{\a\b\g\d}$ as the coefficient at its linear $\theta^{\d}$-dependent term. 
Since we are interested in merely bosonic contributions, we drop the fermionic 
(spinor) components in what follows (except the gravitino-induced bosonic 
torsion).

The chiral density  integration formula reads \cite{ggrw}
\be \lb{chiden}
  \int d^4xd^2\theta\,\ce \Lag =\int d^4x\, e\left\{ 
\Lag_{\rm last} +B\Lag_{\rm first}\right\}~,\ee 
where we have introduced the field components of the covariantly chiral 
superfield Lagrangian $\Lag(x,\theta)$, $ \bar{\de}^{\dt{\a}}\Lag=0$,  as
follows (the vertical bars denote the leading component of a superfield):
\be \lb{comps}
 \left. \Lag\right| =\Lag_{\rm first}(x)~,\qquad
 \left.\de^2\Lag\right|=\Lag_{\rm last}(x)~. \ee

In particular, we have
\be \lb{part}
 \left.\car\right| =\frac{1}{3}\bar{B}=\frac{1}{3}(S+iP)~,\qquad
\left.\de^2\car\right| = \frac{1}{3}\left( R 
-\frac{i}{2}\ve^{abcd}R_{abcd}\right) +\frac{4}{9}\bar{B}B~,\ee
where we have kept the purely imaginary contribution 
$iR_{\rm tor}\equiv\frac{i}{2}\ve^{abcd}R_{abcd}$ because it does not vanish 
in supergravity due to the gravitino- (and matter-) induced torsion $T_{abc}$,
 with  $R_{\rm tor}\propto (\de T +T^2)$. 

\section{Proposal}

A generic supergravity Lagrangian (e.g. representing the supergravitational 
part of the superstring effective action) in superspace is given by
\be \lb{genc}
\Lag = \Lag(\car,\cg,\cw,\ldots) \ee
where the dots stand for arbitrary covariant derivatives of the supergravity 
superfields introduced in Sect.~2. Since the Weyl tensor vanishes for {\it any}
 scale factor in the FRLW metric (\ref{frw}), it is always consistent to take  
$\cw_{\a\b\g}=0$ when discussing the FRLW dynamics (but not its 
perturbations!). Imposing further $\cg_{\a\dt{\a}}=0$ would be too restrictive 
because of the 
Bianchi identities (\ref{bi2}) --- since they would imply $R=const$. Dropping 
the terms with derivatives would generically be inconsistent by the same 
reason.  Nevertheless, we would like to concentrate on the particular 
sector of the theory (\ref{genc}), by ignoring the vector superfield  
$\cg_{\a\dt{\a}}$ and all the derivatives of the superfield $\car$ in 
eq.~(\ref{genc}). Besides having a simplification, we believe that the
 non-scalar arguments in the effective action are not relevant for the dynamics
 of the FRLW scale factor (but they are expected to be relevant e.g. for 
addressing the cosmological singularity \cite{bisw}). So, the effective 
modified supergravity acton we propose is given by
\be
\lb{action}
 S_F = \int d^4xd^2\theta\,\ce F(\car) + {\rm H.c.}
\ee
with some holomorphic function $F(\car)$ presumably generated by strings. 
Besides manifest local $N=1$ supersymmetry, the action (\ref{action}) also
possess the {\it auxiliary freedom} \cite{gat}, since the auxiliary field $B$ 
does not propagate. It distinguishes our action from other possible choices. 
In addition, the action (\ref{action}) automatically gives rise to a spacetime 
torsion.

Most importantly, despite of the apparent presence of higher derivatives, our 
action (\ref{action}) is classically equivalent to the {\it standard} 
supergravity minimally coupled to a chiral `matter' superfield whose chiral  
superpotential is dictated by the chiral function $F$ (see Sect.~8). For 
instance, the purely gravitational part of the action (\ref{action}) in 
components is obtained by eliminating the auxiliary  field $B=S-iP$ via its 
algebraic equation of motion. It results in the modified gravity action having 
the form  
\be \lb{mgrav}
 S_f = \int d^4x \,\sqrt{-g}\, f(R, R_{\rm tor}) \ee
whose function $f$ of scalar curvature $R$ and torsion $R_{\rm tor}$ is 
dictated by the holomorphic function $F$ of the master action (\ref{action}).
The gravitational part of the action (\ref{mgrav}) can be put into the
Brans-Dicke gravity form via a Legendre transform, whereas the Brans-Dicke 
gravity itself is well known to be equivalent to a scalar-tensor gravity
\cite{sot}. Those steps also allow us to ensure the ghost-freedom of our action
 (\ref{action}). When starting from superstrings, the ghost-freedom is 
automatic. The classical equivalence between certain higher-derivative 
supergravities and standard supergravity coupled to matter was observed in 
ref.~\cite{cec} by the use of the superconformal tensor calculus \cite{ku}. 
In the next sections we describe the weak coupling limit of our model and its 
equivalent forms in superspace.

\section{Connection to General Relativity}

Appplying the chiral density formula (\ref{chiden}) to our eq.~(\ref{action}) 
yields the purely bosonic Lagrangian 
\be \lb{expa}
F'(\bar{X}) \left[ \frac{1}{3}R_* +4\bar{X}X \right] +3X F(\bar{X})+{\rm 
H.c.}  \ee
where primes denote differentiation. We have also introduced the notation
\be \lb{not1} X=\frac{1}{3}B~,\qquad R_*=R-iR_{\rm tor}~.\ee

Varying eq.~(\ref{expa}) with respect to the auxiliary fields $X$ and 
$\bar{X}$ gives rise to an algebraic equation on the auxiliary fields,
\be\lb{aux1}
3\bar{F}+X(4\bar{F}'+7F')+4\bar{X}XF'' +\frac{1}{3}F''R_*=0
\ee
and its conjugate,
\be \lb{aux2}
3F+\bar{X}(4F'+7\bar{F}')+4\bar{X}X\bar{F}'' +\frac{1}{3}\bar{F}''\bar{R}_*=0
\ee

First, let's consider the simple special case when 
\be \lb{case1} F''=0~,\qquad {\rm or,~equivalently,} \qquad 
F(\car)=f_0+f_1\car~,\ee
with some complex constants $f_0$ and $f_1$,  where ${\rm Re}f_1<0$.  
Then eq.~(\ref{aux2}) is  easily solved as
\be \lb{sol1}
\bar{X} =\fracmm{-3(f_0+f_1R_*)}{4f_1+7\bar{f}_1} \ee
Substituting the solution (\ref{sol1}) back into the Lagrangian (\ref{expa}) 
yields
\be \lb{gract}
\frac{2}{3}({\rm Re}f_1)R_* - \fracmm{9\abs{f_0}^2}{14({\rm Re}f_1)}
\equiv -\fracmm{1}{2\k^2}R_* -\L = -\fracmm{1}{2\k^2}R(\G+T) -\L \ee
where we have introduced the standard gravitational coupling constant $\k_0=
M^{-1}_{\rm Planck}$ in terms of the (reduced) Planck mass,  the standard 
supergravity connection (i.e. Christoffel symbols $\G$ plus torsion $T$), and 
the cosmological constant $\L$, 
\be \lb{fconst}
  \k= \sqrt{ \fracmm{3}{4\abs{{\rm Re}f_1}}}~~~,\qquad
\L = \fracmm{-9\abs{f_0}^2}{14\abs{{\rm Re}f_1}} \ee
As is clear from the above equations, the cosmological constant in 
supergravity is always negative, as is required by local supersymmetry 
\cite{ggrw}. Since we are not interested in the standard supergravity or 
General Relativity here, we assume that $F''\neq 0$ in what follows.

\section{Superfield Legendre transform}

Our superfield action (\ref{action}) is classically equivalent to another 
action
\be \lb{lmult}
 S_V = \int d^4x d^2\theta\,\ce \left[ \cz\car -V(\cz)\right] + {\rm H.c.}
\ee
where we have introduced the covariantly chiral superfield $\cz$ as the 
 Lagrange multiplier. Varying the action (\ref{lmult}) with respect to 
$\cz$ \footnote{Strictly speaking, one should vary the superfield action with 
respect to an unconstrained \newline ${~~~~}$ pre-potential superfield $U$ 
defined by $\cz = ( \bar{\nabla}^2- 4\car)U$, but it gives rise to the same 
result.} gives back the original action  (\ref{action}) provided that
\be \lb{lt1} 
F(\car) =\car\cz(\car)-V(\cz(\car)) \ee
where the function $\cz(\car)$ is defined by inverting the function
\be \lb{lt2}
\car =V'(\cz) \ee

Equations (\ref{lt1}) and  (\ref{lt2}) define the Legendre transform, and they 
imply further relations,
\be \lb{lt3}
F'(\car)=Z(\car)\qquad {\rm and}\qquad F''(\car)=Z'(\car)
=\fracmm{1}{V''(\cz(\car))}  \ee
where $V''=d^2V/d\cz^2$. The second formula (\ref{lt3}) is the duality relation
 between the supergravitational function $F$ and the chiral superpotential $V$.

\section{Modified gravity from supergravity}

The field equations (\ref{aux1}) and (\ref{aux2}) are easily solved for
$R_*=R-iR_{\rm tor}$~~,
\be \lb{aux3}
R = {\rm Re}\,R_*=R(X,\bar{X})=
- {\rm Re}\,\fracmm{9\bar{F}+3X(4\bar{F}'+7F')}{F''}-12\bar{X}X \ee
and
\be \lb{aux4}
R_{\rm tor}=R_{\rm tor}(X,\bar{X})={\rm Im}\,
\fracmm{9\bar{F}+3X(4\bar{F}'+7F')}{F''} \ee
Inverting those functions and substituting the result back into the component
action (\ref{expa}) gives rise to the modified gravity action (\ref{mgrav}) 
with 
\be \lb{mgrav2}
f(R,R_{\rm tor}) = \left.
F'(\bar{X}) \left[ \frac{1}{3}R_* +4\bar{X}X \right] +3X F(\bar{X})
+{\rm H.c.} \right|_{X=X(R,R_{\rm tor})} \ee

\section{Weyl transform in components}

Let's take $R_{\rm tor}=0$ in eq.~(\ref{mgrav}) for even more simplicity, and 
rescale the
function $f(R)$ to $(-1/2\k^2) f(R)$ with the gravitational coupling constant 
$\k$. Then the action (2.8) is classically equivalent to 
\be \lb{rmgr}
S_A = \fracmm{-1}{2\k^2}\int d^4x\,\sqrt{-g}\,\left\{ AR-V(A)\right\} \ee
where the real scalar $A(x)$ is related to the scalar curvature $R$ by the 
Legendre transform,
\be \lb{clt} R=V'(A) \qquad{\rm and}\qquad f(R)=RA(R)-V(A(R)) \ee 

A Weyl transformation of the metric,
\be \lb{weylm}
g_{\m\n}(x)\to \exp \left[ \fracmm{2\k\f(x)}{\sqrt{6}} \right] g_{\m\n}(x)\ee  
with an arbitrary parameter $\f(x)$, yields
\be \lb{weylr}
\sqrt{-g}\,R \to \sqrt{-g}\, \exp \left[ \fracmm{2\k\f(x)}{\sqrt{6}} \right]
\left\{ R -\sqrt{\fracmm{6}{-g}}\pa_{\m}\left(g^{\m\n}\pa_{\n}\f\right)\k
-\k^2g^{\m\n}\pa_{\m}\f\pa_{\n}\f\right\} \ee
Hence, when choosing 
\be \lb{ch1}
A(\k\f) = \exp \left[ \fracmm{-2\k\f(x)}{\sqrt{6}} \right]  \ee
we can rewrite the Lagrangian in eq.~(\ref{rmgr}), up to a total derivative, to
the form
\be \lb{stgr}
S_{\f} =  \int d^4x\, \sqrt{-g}\left\{ \fracmm{-R}{2\k^2} 
+\fracmm{1}{2}g^{\m\n}\pa_{\m}\f\pa_{\n}\f + \fracmm{1}{2\k^2}
\exp \left[ \fracmm{4\k\f(x)}{\sqrt{6}}\right] V(A(\k\f)) \right\} \ee
in terms of the physical (and canonically normalized) scalar field $\f(x)$.

This procedure is well known in the $f(R)$ gravity theories with {\it ad hoc} 
functions $f(R)$ --- see e.g. ref.~\cite{sot} for a recent review. The $f(R)$
 modification of Einstein gravity is the alternative to the dark energy for
 explaining the present acceleration of the Universe in the IR limit (or 
large distances).~\footnote{It should be $f''(R)>0$ in order to avoid the 
so-called Dolgov-Kawasaki instability \cite{dkaw}.} Our motivation for the 
$F(\car)$ supergravity comes from the UV limit (or small distances) to be 
described by a UV-complete theory of superstrings, but due to the universal 
nature of superstrings, the same effective gravity may still be valid in the 
IR limit (after some renormalization and supersymmetry breaking). Accordingly, 
we would like to interpret the scalar field $\f$ as an {\it inflaton} in the 
early Universe and as the {\it quintessence} in the present Universe, with 
a scalar potential
\be \lb{infp}
W(\f)= - \fracmm{1}{2\k^2}\exp \left[ \fracmm{4\k\f(x)}{\sqrt{6}}\right] 
V(A(\k\f))\ee
It is worth mentioning that the effective gravitational coupling constant $\k$
here may be different from its naive value $\k_0$ in the IR-limit, due to 
possible renormalization effects. As regards a space-time torsion in the 
$f(R)$-gravity (though unrelated to spin fields), see e.g. ref.~\cite{frtor}. 

\section{Super-Weyl transform in superspace}

A super-Weyl transform of the superfeld acton (\ref{lmult}) can be done
entirely in superspace, i.e. with manifest local N=1 supersymmetry. In terms
of components, the super-Weyl transform amounts to a Weyl transform, a chiral 
rotation and a (superconformal) $S$-supersymmetry transformation \cite{ht}.
 The chiral density superfield $\ce$ is just the chiral compensator of the 
super-Weyl transformations
\be \lb{swt}
\ce \to e^{3\k\F} \ce~, \ee
whose parameter $\F$ is an arbitrary covariantly chiral superfield,
$\bar{\de}_{\dt{\a}}\F=0$. Under the transformation (\ref{swt}) the 
covariantly chiral superfield $\car$ transforms as 
\be \lb{rwlaw}
\car \to e^{-2\k\F}\left( \car - \fracm{1}{4}\bar{\nabla}^2\right)
e^{\k\bar{\F}}
\ee
When choosing the super-Weyl chiral superfield parameter to obey 
\be \lb{ch2} \k\F = \fracmm{1}{\x}\ln\cz~, \ee
the super-Weyl transform of the acton (\ref{lmult}) gives rise to the
classically equivalent action~\footnote{We have rescaled the action by a factor
of $-3/\k^2$ and introduced an arbitrary (real) number \newline ${~~~~~}$
$\x$,  while keeping the normalization of $\F$ arbitrary.}  
\be \lb{chimat}
 S_{\F} =-\fracmm{3}{\k^2} \int d^4x d^2\theta\, \ce \left\{ 
 e^{(1+\x)\k\F}(\car -\frac{1}{4}\bar{\nabla}^2)e^{\k\bar{\F}}- 
e^{3\k\F}V(e^{\x\k\F}) \right\}  +{\rm H.c.} \ee
or
\be \lb{chimat2}
\eqalign{
S_{\F} = &  -\fracmm{3}{\k^2} \int d^4x d^4\theta\, E^{-1} e^{\k(\F+\bar{\F})}
\left( e^{\x\k\F}+e^{\x\k\bar{\F}}\right) \cr
& +\left[ \fracmm{3}{\k^2}  \int d^4x d^2\theta\, \ce e^{3\k\F}V(e^{\x\k\F})
+{\rm H.c.} \right]~~, }
\ee
where we have introduced the full superspace supergravity supervielbein 
$E^{-1}$ \cite{ggrw}.

Equation (\ref{chimat2}) has the standard form of a chiral superfield action
coupled to supergravity, in terms of a K\"ahler potential $K(\F,\bar{\F})$ and 
a chiral superpotential $W$, with
\be \lb{stand}
K(\F,\bar{\F})= -\fracmm{3}{\k^2}e^{\k\F +\k\bar{\F}}
\left( e^{\x\k\F}+e^{\x\k\bar{\F}}\right)~,\qquad
W(\F)= \fracmm{3}{\k^2} e^{3\k\F}V(e^{\x\k\F}) \ee
Therefore, the associated scalar potential is given by the standard formula 
\cite{crem}
\be \lb{crem}
 \cv (\f,\bar{\f}) =\left. e^K \left\{ \abs{\frac{\pa W}{\pa\F}
+\frac{\pa K}{\pa\F}W}^2-3\k^2\abs{W}^2\right\} \right|~, \ee
where all superfields are restricted to their leading field components,
$\left.\F\right|=\f(x)$.

Equations (\ref{lt1}), (\ref{lt2}), (\ref{stand}) and (\ref{crem}) give the 
algebraic relations between a function $F$ in our supergravity action 
(\ref{action}) and a scalar potential $\cv$ of the classically equivalent 
scalar-tensor supergravity (\ref{chimat2}). In particular, eq.~(\ref{crem}) 
can used for embedding inflation into supergravity. Now it can be promoted 
further, by embedding inflation into the `purely geometrical' modified 
supergravity (\ref{action}) that determines a K\"ahler potential and a chiral 
superpotential of the inflaton superfield  in terms of a single holomorphic 
function $F$.

\section{Discussion}

A possibility to achieve inflation by modifying Einstein equations with the 
2nd-order curvature terms (representing the gravitational anomalies of the
matter fields) was discovered a long time ago \cite{staro}. A similar 
mechanism exist in the four-dimensional supergravity, with inflation generated 
by the $\car^2$-term originating from the one-loop K\"ahler anomaly 
\cite{carov}. The instabilities in the scenarios based on the 2nd-order 
curvature terms against adding the higher order scalar curvature terms were 
discussed in ref.~\cite{maeda} within perturbation theory.~\footnote{Those 
instabilities can be suppressed against the supersymmetric ${\car}^2$ terms 
\cite{carov}.} The inflationary solutions generated by the purely 4th-order 
terms in the curvatures, in the effective supergravity action generated by 
superstrings were found in ref.~\cite{iihk}. Their stability and the scale 
factor duality invariance were also investigated \cite{iihk}. In this Letter 
we emphasize the significance of the full non-perturbative structure of a 
holomorphic function $F(\car)$ ({\sf cf.} the Born-Infeld-type supergravity 
\cite{gk}). The higher-order curvature terms  ${\car}^n$ are also generated by 
radiative corrections in supergravity \cite{carov} though, unlike superstrings,
they  cannot be consistent because of the non-renormalizabilty of supergravity.

In General Relativity, only the spin-$2$ part of a metric is dynamical. The 
dynamical generation of a massive scalar field is known to occur already in the
 presence of the quadratic curvature terms \cite{buch}, namely, out of the 
spin-$0$ part of the metric. In supergravity, as was shown in the preceeding 
section, the whole chiral scalar superfield becomes dynamical,  while it can be
 identified with a super-Weyl compensator --- see eq.~(\ref{ch2}). In 
superstring theory, the superspin-$0$ part of the supervielbein is given by a 
chiral scalar superfield, whose leading complex component represents a 
{\it dilaton-axion} field, $\left.\f\right|=\vf(x)+iB(x)$. Hence, we identify 
$\vf(x)$ with a superstring dilaton, and $B$ with a superstring $B$-field (or 
axion).~\footnote{Since the dilaton is already present in the spectrum of 
superstrings, the coefficient at the \newline ${~~~~}$ kinetic term in 
eq.~(\ref{stgr}) should be modified.} As is well known in string theory 
\cite{bbs}, the dilaton field controls the superstring loops and (D-brane)
instantons, which may be the source of the function $F(\car)$. The $B$-field 
is the source of the non-minimal space-time torsion in string theory 
\cite{myb}. 

Unfortunately, the string theory technology at present does not allow us to 
compute the function $F(\car)$ in eq.~(\ref{action}). It is mainly because of 
the {\it on-shell} nature of the known string theory that, in principle, can 
unambiguously fix only the $\cw_{\a\b\g}$-dependence of the gravitational 
effective action \cite{myb}. However, its $\car$-dependence can be fixed by 
some additional (off-shell) physical requirements such as no-ghosts, stability 
and the scale-factor self-duality \cite{iihk}, or by going to the IR-limit 
(weak gravity). For instance, by using hameleon effect, it was demonstarted in 
ref.~\cite{starnew} that  the function $f(R)$ with
\be\lb{stn} f(R) =R + \l R_0 \left[ \fracmm{1}{\left(1+
\fracmm{R^2}{R^2_0}\right)^n} -1\right] \ee
with some parameters $R_0\sim H^2_0$, $\l>0$ and $n>0$, is fully consistent 
with {\it all\/} Solar System observations. Of course, there are many other 
acceptable choices \cite{more}, e.g., by the {\it reconstruction} of the 
function $f(R)$ from a desired (given) scale factor $a(t)$ via the 
gravitational equations of motion with 
\be \lb{scurv} R = -6 \left[ \fracmm{\ddt{a}}{a}+\left(
\fracmm{\dt{a}}{a}\right)^2+\fracmm{k}{a^2}\right] \ee
The superfield extension of eq.~(\ref{scurv}) is given by a 
superconformally-flat superspace with 
\be \lb{sflat} \car = -\fracmm{1}{4}e^{-2\k\F}\bar{\nabla}^2e^{\k\bar{\F}}\ee
so that the function $F(\car)$ can also be reconstructed via the equations of 
motion from a desired history $a(t)$.

\end{document}